\documentclass[11pt,twoside]{article}

\usepackage{asp2006}
\usepackage{epsf}
\usepackage{lscape}

\begin{document}
\title{Differential Evolution of Lyman Break Galaxies from $z$=5 to 3}   
\author{Ikuru Iwata\altaffilmark{1}, Kouji Ohta\altaffilmark{2}, Marcin Sawicki\altaffilmark{3}, Naoyuki Tamura\altaffilmark{4}, Masayuki Akiyama\altaffilmark{4}, Kentaro Aoki\altaffilmark{4}}   
\altaffiltext{1}{Okayama Astrophysical Observatory, National Astronomical Observatory of Japan}    
\altaffiltext{2}{Department of Astronomy, Kyoto University, Japan}
\altaffiltext{3}{Department of Astronomy and Physics, St. Mary's University, Canada}
\altaffiltext{4}{Subaru Telescope, National Astronomical Observatory of Japan} 

\begin{abstract} 
We briefly summarize our findings from the unbiased surveys for $z$$\sim$5 LBGs 
based on Subaru/Suprime-Cam and follow-up optical spectroscopy.
\end{abstract}


We have been conducting systematic surveys of Lyman break galaxies (LBGs) 
at $z\sim 5$ (the cosmic age $\sim$1.2Gyr). 
It is the highest redshift we can apply two-color selection method 
for LBG candidates with well-defined optical filters. 
Our $z$$\sim$5 LBG sample is based on surveys for the 
two independent blank fields (the region including the Hubble Deep 
Field - North and the J0053+1234) obtained with the Subaru / Suprime-Cam.
The total effective area after masking  
bright objects is 1,300 arcmin$^2$, and deep $V$, $I_c$ and $z'$-band 
imaging enabled us to securely select $V$-dropout objects down to 
$z'_\mathrm{AB}$=26.5 mag (for the HDF-N region) or 25.5 mag (for the 
J0053+1234 region). The number of $z\sim5$ LBG candidates in our sample is 
850. It should be emphasized that the area coverage of our survey 
is more than 100 times wider than 
the ACS field of the Hubble Ultra Deep Field and more than 4 times 
wider than the total area covered by the GOODS, and this wide field 
coverage has a crucial importance for reliable determination of the 
abundance of luminous objects. 
The redshifts of a number of our LBG candidates 
have been spectroscopically determined, 
and the validity of our color selection criteria have been confirmed 
(\citeauthor{ando04} \citeyear{ando04}; \citeyear{ando07}).

\begin{figure}[!ht]
\plottwo{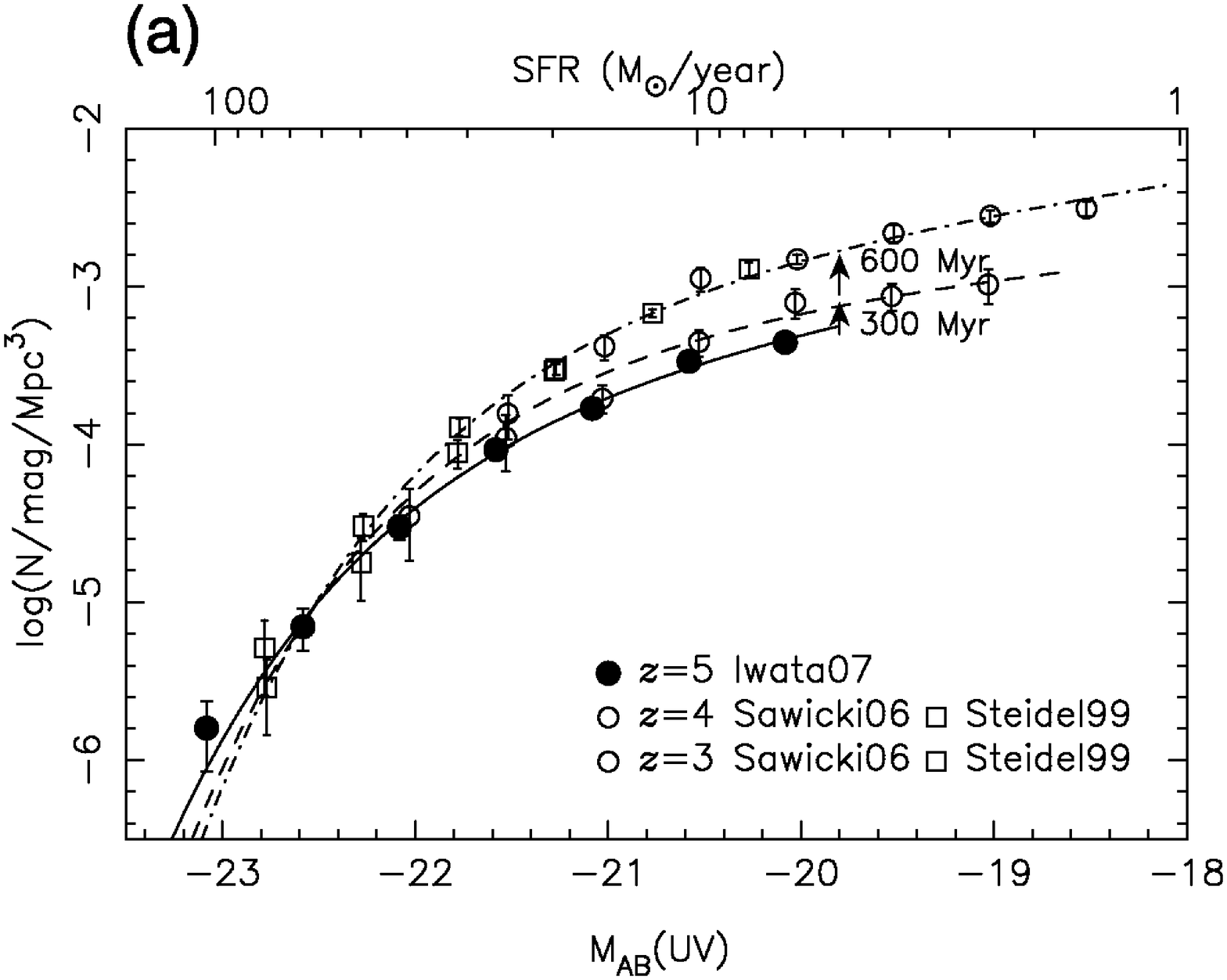}{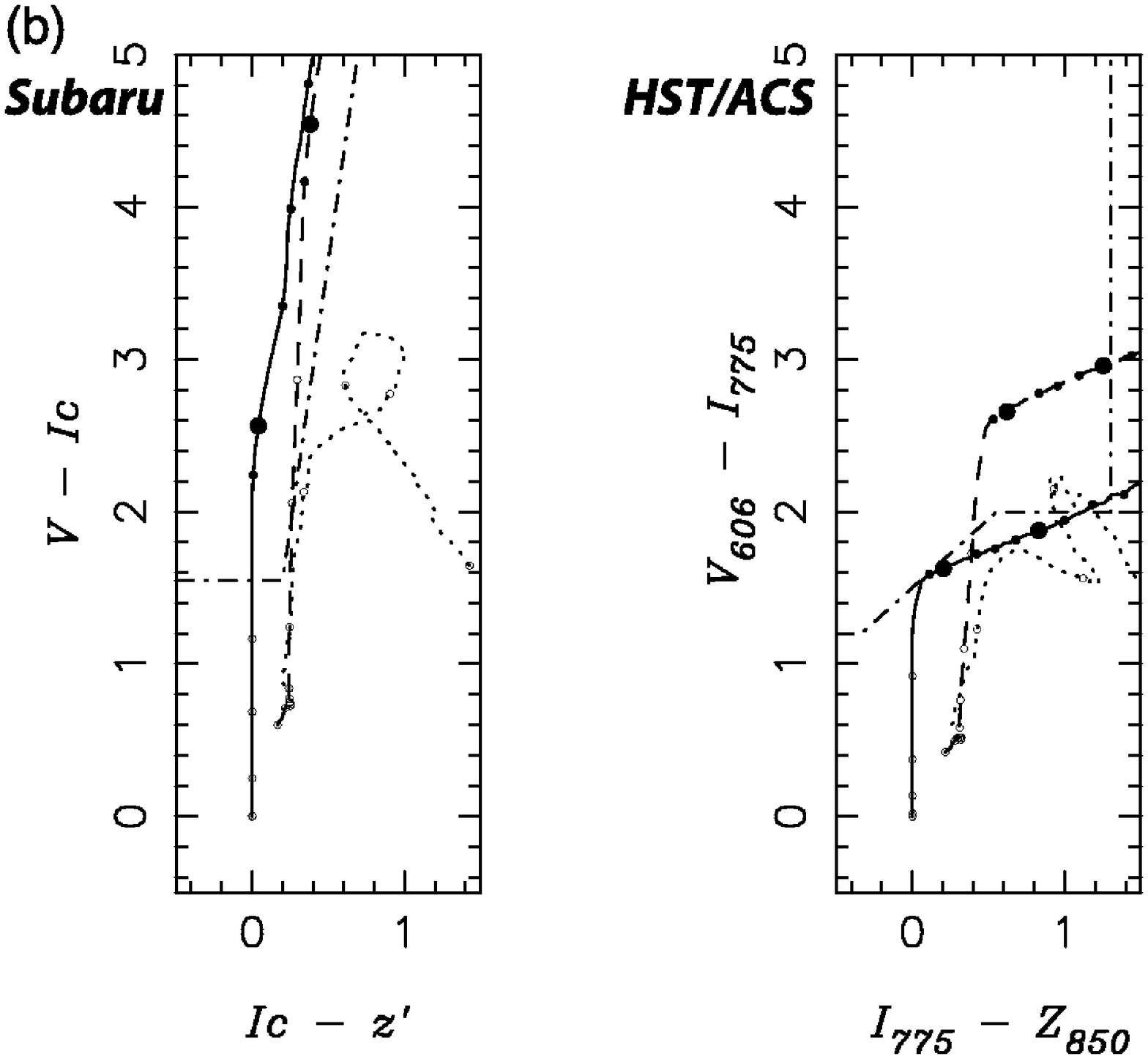}
\caption{(a) UV luminosity function of LBGs at $z\sim5$ (solid), 
4 (dashed) and 3 (dot-dashed). (b) Comparison of two-color diagrams used for selection of $z\sim5$ LBGs. 
Left: a case for Subaru/Suprime-Cam filter set used in Iwata et al. (2007). Right: 
a case for HST/ACS filter set.}
\label{fig:LF}
\end{figure}

In figure~\ref{fig:LF}(a) we show the UV luminosity function (LF) of 
LBGs at $z \sim 5$ derived from our sample with filled circles and a solid line 
\citep{I07}. 
In this figure we also 
show the UVLF of LBGs at $z\sim 4$ and 3 based on the very deep survey 
\citep[Keck Deep Fields;][]{sawicki06}. 
We found that in the luminous end 
of the UV LF there is no significant evolution from $z\sim5$ to 
3 ($\approx$1 Gyr), while in the fainter part, the gradual 
increase of number density is observed. 
This clear contrast 
in the UV LF suggests that the evolution of the LBGs is 
{\it differential}, depending on UV luminosity.

In spectroscopic follow-up observations we also found that 
equivalent widths of Ly-$\alpha$ emission for star-forming galaxies 
at $z=5$--6 show a strong dependence on UV luminosity \citep{ando06}: 
UV luminous objects have weak or no Ly-$\alpha$ emission, suggesting that 
they are either in relatively dusty environment or are enshrouded 
by massive HI gas haloes. 
We suggest that the evolution of star-forming 
galaxies in the first 2 Gyr of the universe could be well described 
with the biased evolution scenario:  
a galaxy population hosted by massive dark haloes start active 
star formation preferentially at early time of the universe, 
while less massive galaxies increase their number density later. 
To understand the origin of this differential evolution would 
be an important clue to clarify the star-formation process 
in the early universe.

However, we should note that there is a controversy on the 
evolution of the UV LF at $z>3$. 
Some authors claim a rapid evolution in the 
bright end (e.g., \citeauthor{yoshida06}\citeyear{yoshida06}; 
\citeauthor{bouwens07}\citeyear{bouwens07}). 
Because the filter sets used in these studies are 
different from each other, a great care should be paid for the differences of 
color selection criteria and selected samples. 
As an example we show in figure \ref{fig:LF}(b) two color selection criteria 
used in our study and those in \citet{oesch07} who used HST/ACS data 
to select LBGs at $z\sim5$, as well as model color tracks (
flat $f_\nu$ at rest-frame UV wavelengths, 
dust attenuation $E(B-V)$=0.4 and template elliptical galaxy at $z<2$).
Due to the broadness of $V_{\mathrm 606}$ filter, color tracks in ACS filter 
set show a break at $z$$\approx$5, and it suggests the difficulty in proper selection 
of $z\sim5$ LBG candidates with HST/ACS filters. 
Understanding differences of samples in various studies would be required 
to reconcile the different views on the evolution of LBGs at $z>3$.





\end{document}